\documentclass[journal=nalefd,manuscript=letter]{achemso}
\setkeys{acs}{articletitle = true}
\setkeys{acs}{etalmode = truncate}
\setkeys{acs}{maxauthors = 20}
\usepackage[version=3]{mhchem} 
\usepackage{amssymb}
\usepackage{amsmath,xcolor,bbm}

\DeclareFontFamily{OMS}{oasy}{\skewchar\font48 }
\DeclareFontShape{OMS}{oasy}{m}{n}{%
         <-5.5> oasy5     <5.5-6.5> oasy6
      <6.5-7.5> oasy7     <7.5-8.5> oasy8
      <8.5-9.5> oasy9     <9.5->  oasy10
      }{}
\DeclareFontShape{OMS}{oasy}{b}{n}{%
       <-6> oabsy5
      <6-8> oabsy7
      <8->  oabsy10
      }{}
\DeclareSymbolFont{oasy}{OMS}{oasy}{m}{n}
\SetSymbolFont{oasy}{bold}{OMS}{oasy}{b}{n}

\DeclareMathSymbol{\smallleftarrow}     {\mathrel}{oasy}{"20}
\DeclareMathSymbol{\smallrightarrow}    {\mathrel}{oasy}{"21}
\DeclareMathSymbol{\smallleftrightarrow}{\mathrel}{oasy}{"24}

   \let\vec\mathbf

\author{Andrew W. Rossi}
\altaffiliation{These authors contributed equally}
\affiliation
{Department of Chemistry, University of Washington, Seattle, WA, 98195}
\author{Marc R. Bourgeois}
\altaffiliation{These authors contributed equally}
\affiliation
{Department of Chemistry, University of Washington, Seattle, WA, 98195}
\author{Caleb Walton}
\affiliation
{Department of Chemistry, University of Washington, Seattle, WA, 98195}
\author{David J. Masiello}
\email{masiello@uw.edu}
\affiliation
{Department of Chemistry, University of Washington, Seattle, WA, 98195}

\title[]
{Probing the Polarization of Low-Energy Excitations in 2D Materials from Atomic Crystals to Nanophotonic Arrays using Momentum-Resolved Electron Energy Loss Spectroscopy}


\begin{document}





\begin{tocentry}
\centering
\includegraphics{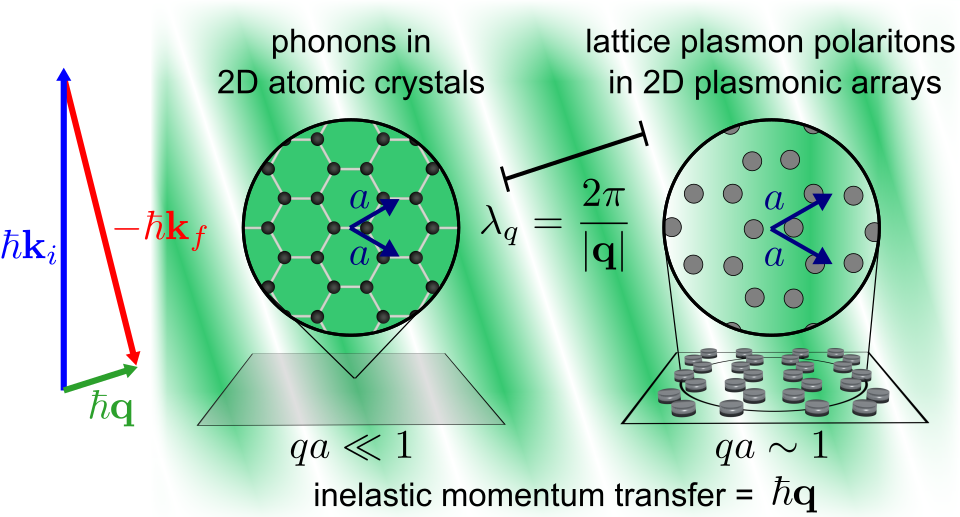}
\end{tocentry}


\begin{abstract}
Spectroscopies utilizing free electron beams as probes offer detailed information on the reciprocal-space excitations of 2D materials such as graphene and transition metal dichalcogenide monolayers. Yet, despite the attention paid to such quantum materials, less consideration has been given to the electron-beam characterization of 2D periodic nanostructures such as photonic crystals, metasurfaces, and plasmon arrays, which can exhibit the same lattice and excitation symmetries as their atomic analogs albeit at drastically different length, momentum, and energy scales. Due to their lack of covalent bonding and  influence of retarded electromagnetic interactions, important physical distinctions arise that complicate interpretation of scattering signals. Here we present a fully-retarded theoretical framework for describing the inelastic scattering of wide field electron beams from 2D materials and apply it to investigate the complementarity in sample excitation information gained in the measurement of a honeycomb plasmon array versus angle-resolved optical spectroscopy in comparison to single monolayer graphene. 
\end{abstract}

Historically, techniques employing low energy electrons, such as low energy electron diffraction (LEED), scanning tunneling microscopy (STM), high resolution electron energy-loss spectroscopy (HREELS), and photoemission spectroscopy (PES), have played a critical role in understanding a range of material properties, from local atomic structure \cite{hamers1989atomic,saldin1990holographic,bauer1994low,wolkow1992direct} to dispersion relations of materials hosting collective excitations, such as phonons \cite{siebentritt1997surface,aizawa1990anomalous} and plasmons \cite{moresco1999plasmon, petek1997femtosecond, diaconescu2007low, bostwick2007quasiparticle}. The isolation of graphene \cite{novoselov2004electric} and other 2D atomic crystals \cite{novoselov2005two, naguib2011two, liu2014phosphorene} has furthered the need for state-of-the-art characterization techniques where the excitonic \cite{hong2020probing,mak2016photonics,ross2013electrical}, phononic \cite{maultzsch2004phonon,mohr2007phonon,politano2015emergence,politano2017spectroscopic,li2023direct,senga2019position}, and plasmonic \cite{kramberger2008linear,chen2012optical,ni2018fundamental,zhou2012atomically} excitations in these materials underlie important applications in optoelectronic devices \cite{chen2012optical,liu2011graphene,bonaccorso2010graphene,mak2016photonics} and quantum information technologies \cite{liu20192d}. Due to its atom scale spatial resolution and ability to measure broadband spectral responses, electron energy-loss spectroscopy (EELS) performed inside a scanning transmission electron microscope (STEM) has played an important role in characterizing such 2D materials at their native response scales \cite{krivanek2014vibrational,dwyer2016electron,lagos2017mapping,hage2018nanoscale,zhou2012atomically}. However, the high spatial resolution offered by STEM-EELS relies on momentum space integration, limiting its use as a probe of reciprocal-space excitations. By increasing the incident beam width, parallel beam momentum-resolved EELS ($q$-EELS) in a STEM or wide field $q$-EELS in a TEM overcome this challenge, offering sufficient momentum resolution to characterize the dispersive responses of graphene \cite{senga2019position,kramberger2008linear} and other 2D materials \cite{senga2019position,hong2020probing,hage2018nanoscale}. Similarly, HREELS, adapted from its original uses in surface science, has demonstrated the ability to retrieve detailed excitation information of 2D materials with even higher momentum and energy resolution \cite{politano2015emergence,politano2017spectroscopic,li2023direct,politano2021fate}. 

Like 2D quantum materials with atomic scale periodicities, 2D periodic nanophotonic structures, such as photonic crystals, metasurfaces, and plasmon arrays, have been under intense study owing to their facilely tunable optical properties and role in the formation of new hybrid light-matter states via interaction with excitonic media in both weak and strong coupling regimes. Actively tunable array lasers \cite{yang2015real,Torma_Kpoint_2019,juarez2022m,hirose2014watt} and exciton polariton structure, dynamics \cite{liu2020generation, bourgeois2022nanometer,doi:10.1021/acs.nanolett.0c01624,Troma_Strong_2014,shi2014spatial,baranov2018novel}, and room temperature condensation \cite{Torma_BEC_2019, berghuis2023room} represent examples that exploit the energy-momentum dispersion arising from the discrete space translational symmetry of the lattice \cite{cherqui2019plasmonic, Manjavacas2018} to create coherent optical states or simulate complex quantum many-body phenomena. Due to their ability to support directional and vector-polarized lasing emission, bound states in the continuum (BIC) realized in periodic nanophotonic structures have also been in the spotlight \cite{zhen2014topological,azzam2018formation,liang2020bound, guan2020quantum, liang2021hybrid,ha2018directional, mohamed2022controlling}. Measurements of these phenomena necessitate the use of characterization tools with simultaneously high energy, momentum, and polarization resolution. While optical-based probes have been successful in the characterization of such materials, the use of $q$-EELS has been limited to the study of photonic density of states in the vicinity of plasmonic thin film architectures \cite{shekhar2017momentum} and crystal defects \cite{saito2019emergence}, despite its ability to probe sample excitations with high energy and momentum resolution that are inaccessible to light. Like STEM-EELS, a powerful tool for mapping single nanoparticle excitations \cite{batson1982new,nelayah2007mapping,de2010optical,cherqui2016characterizing}, momentum-resolved EEL measurements using wide field electron beams have strong potential in the characterization of periodically structured nanophotonic materials \cite{polman2019electron}, as exemplified by the routine use of $q$-EELS and HREELS to interrogate the dispersive features of 2D atomic crystals.

Here we present a common theoretical framework for the interpretation of wide field $q$-EELS signals in the measurement of the broadband reciprocal-space excitations in 2D crystalline materials spanning from atomic to nanophotonic dimension and expose the unexplored ability of $q$-EELS to probe the 3D polarization-resolved responses of plasmonic arrays. Reflection scattering processes typical of HREELS are also equally treated from the same formalism. While well understood as a probe of responses intrinsic to solid state 2D materials with {\AA}ngstrom-scale bond lengths, the $\gtrsim100$ nm periodicity scale of 2D nanophotonic materials such as photonic crystals, metasurfaces, and plasmon arrays present a distinctly new physical regime where electromagnetic retardation effects prevail and companion wide field inelastic electron scattering observables must be interpreted accordingly. Through the lens of expanded selection rules appropriate to fully retarded light-matter interactions, we elucidate the conditions under which $q$-EELS can be adapted for measuring the broadband excitations of nanophotonic 2D materials with emphasis placed upon the evolution of the 3D selection rules for scattering processes both within and beyond the first Brillouin zone (BZ), the excitation of optically dark transitions on and off of the light cone, and role of electron speed and collection geometry (i.e., transmission versus reflection). Throughout, contrast and comparison between the IR vibrational excitations of graphene and the optical-frequency excitations of a plasmonic honeycomb lattice as encoded in the inelastic electron scattering signal are made to illustrate the utility of $q$-EELS in probing 2D materials excitations spanning broadly across disparate spatial, momentum, and energy scales.

\begin{figure}
\centering
\includegraphics{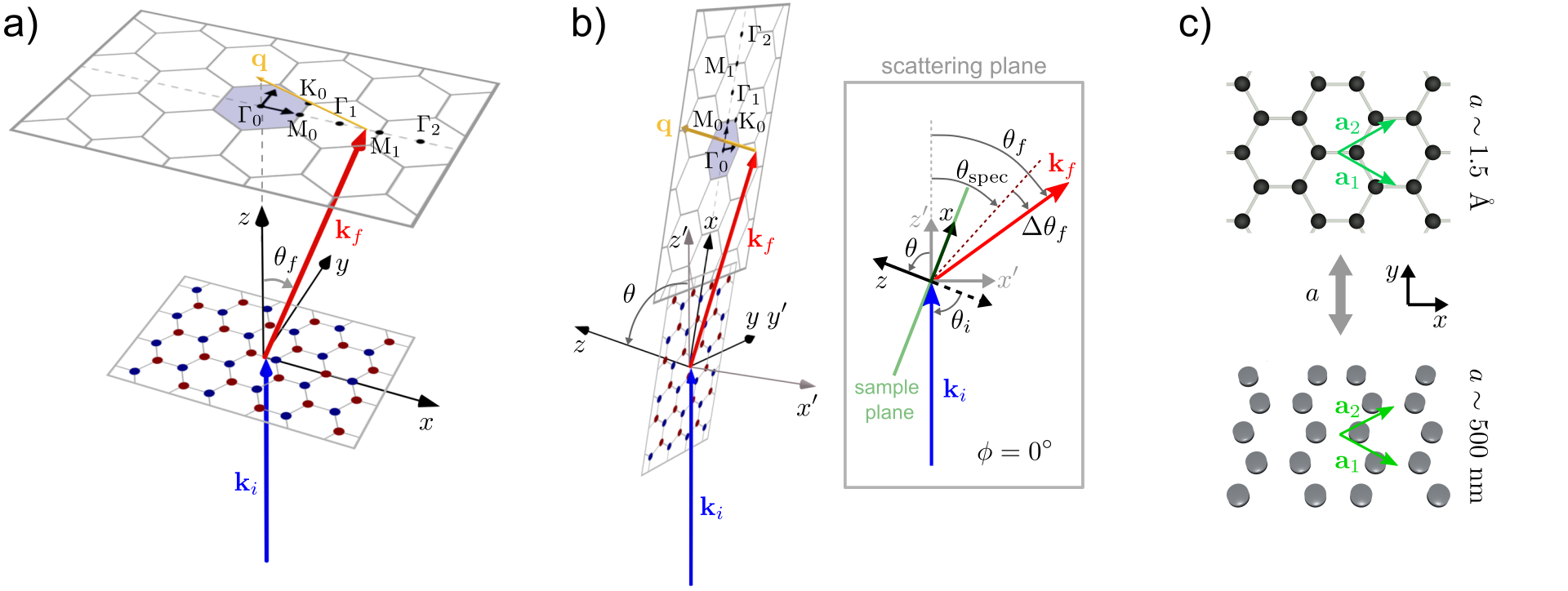}
\caption{Probing reciprocal-space excitations in 2D material systems with $q$-EELS. a) Scheme of $q$-EELS measurement under forward scattering conditions. Momentum transfers between initial $\mathbf{k}_i$ (blue) and final $\mathbf{k}_f$ (red) free electron states are dictated by the scattering angle $\theta_f$ and recoil momentum $\hbar{\bf q}$ (yellow) with projection ${\bf q}_\parallel=({\bf q}\cdot\hat{\bf x})\hat{\bf x}$ oriented along the $\Gamma-\textrm{M}$ direction of the sample. b) Scheme of $q$-EELS measurement under reflection scattering conditions, otherwise sharing the same kinematical parameters as in transmission. The inset defines the $xz$ scattering plane relevant to both transmission and reflection geometries. The detection angle is $\theta_f$ in transmission and $\Delta\theta_f=\theta_f-\theta_{\textrm{spec}}$ in reflection, where $\theta_{\textrm{spec}}$ is the specular angle. Lab (primed) and sample (unprimed) coordinates are distinguished, and the $xz$-scattering plane is always perpendicular to the 2D sample $xy$-plane. Beginning with coincident lab and sample axes as in (a), where only the sample axes are explicitly labeled, and collecting momentum recoils within the $x'z'$-plane, an arbitrary planar scattering configuration can be reached by (i) rotating the sample by angle $\phi$ about the $z'$-axis, followed by (ii) rotation by angle $\theta$ about the lab $y'$-axis. c) Graphene (upper) and a plasmonic array (lower) are examples of 2D periodic honeycomb lattices structured on the atomic and photonic length scales, respectively. $a$ is the magnitude of the primitive vectors $\mathbf{a}_1$ and $\mathbf{a}_2$ spanning the real space array. In either case, point $\textrm{M}_0$ is located at $\mathbf{q}_{\parallel} = (2\pi/\sqrt{3}a, 0)$.}
\label{F1}
\end{figure}

The schemes in Figs. \ref{F1}a,b show a pair of generic wide field inelastic electron scattering events whereby an incoming free electron plane wave with wave vector $\mathbf{k}_i$ (blue) scatters to an outgoing plane wave with wave vector $\mathbf{k}_f$ (red) via interaction with a 2D periodic material sample. The linear momentum $\hbar \mathbf{q} = \hbar (\mathbf{k}_i - \mathbf{k}_f)$ lost by the probing electron is transferred into excited sample states, inducing the transition $|0\rangle\to|n\rangle$. Experimental conditions, such as detection of transmitted (Fig. \ref{F1}a) versus reflected (Fig. \ref{F1}b) electrons as well as the specific incident and collection angles involved, fix $\mathbf{q}$ (yellow), while the projection $\mathbf{q}_\parallel$ of $\mathbf{q}$ onto the $xy$ sample plane can be manipulated by rotating the sample. Unless stated otherwise, $\phi=0^{\circ}$ such that transverse recoils ${\bf q}\cdot\hat{\bf x}$ in the $xz$ scattering plane are connected with reciprocal space excitations in the 2D-periodic samples with Bloch vectors $(\mathbf{K}\cdot\hat{\bf x})\hat{\bf x}.$

As representative 2D-periodic samples that support reciprocal space excitations in the low-loss regime, we consider graphene and a plasmonic honeycomb array (Fig. \ref{F1}c), which support phonons and lattice plasmon polaritons (LPPs), respectively. The 2D atomic (nanoparticle) positions $\mathbf{x}_{\mathbf{n}\kappa} = \mathbf{x}_{\mathbf{n}}+{\bf r}_\kappa=n_1 \mathbf{a}_1 + n_2 \mathbf{a}_2+{\bf r}_\kappa$ in the graphene (plasmonic array) case are situated on a honeycomb lattice, which is a hexagonal Bravais lattice described by primitive lattice vectors $\mathbf{a}_{1}$ and $\mathbf{a}_{2}$, with $|\mathbf{a}_1| = |\mathbf{a}_2| = a$ and $s=2$ sites per unit cell labeled by $\kappa$. These sites are colored red and blue in Figs. 1a,b. While both honeycomb arrays share common real space and reciprocal space symmetry, the characteristic length scale is $a=1.42$ {\AA} for graphene, and $a=460$ nm for the plasmonic array.

The Hamiltonian involving either class of 2D material can be partitioned as $\hat{H} = \hat{{H}}_0 + \hat{{H}}_\textrm{int}$, where $\hat{{H}}_0 = \hat{{H}}_s + \hat{{H}}_\textrm{el}$ is the sum of sample $\hat{{H}}_s$ and free electron $\hat{{H}}_\textrm{el}$ components, and $\hat{{H}}_\textrm{int}$ is the interaction Hamiltonian. Beginning from minimal coupling and working in the generalized Coulomb gauge \cite{glauber1991quantum}, the interaction Hamiltonian is $\hat{{H}}_{\textrm{int}} = (e/2mc)[ \hat{\mathbf{A}} \cdot \hat{\mathbf{p}} + \hat{\mathbf{p}} \cdot \hat{\mathbf{A}}]$, where $m$ and $-e$ are the electron mass and charge, and $\hat{\mathbf{A}}$ and $\hat{\mathbf{p}}$ are the quantum mechanical operators associated with the vector potential of the target and the linear momentum of the free electron probe, respectively. The wide field double differential cross section (DDCS) with fully-retarded probe-sample interactions becomes \cite{doi:10.1126/sciadv.adj6038,nixon2023inelastic}
\begin{equation}
    \frac{\partial^2 \sigma}{\partial E_{if} \partial \Omega_f} = \Big(\frac{m L^3}{\pi\hbar^2}\Big)^2 \bigg( \frac{ k_{f}}{k_i} \bigg) \int d\omega \, d\mathbf{x} d\mathbf{x}' \, \textrm{Im}\Big[ \mathbf{J}_{fi}^{*}(\mathbf{x}) \cdot \bar{\mathbf{G}}(\mathbf{x}, \mathbf{x}', \omega) \cdot \mathbf{J}_{fi}(\mathbf{x}') \Big]\delta(\omega-\varepsilon_{if}),
    \label{DDCS_final_form}
\end{equation}
where $L$ is the box quantization length, $E_{if}=\hbar \varepsilon_{if} = \hbar\varepsilon_i - \hbar\varepsilon_f $ is the loss energy, and $\hbar \varepsilon_{i/f} = \gamma_{i/f}mc^2$ with initial/final Lorentz contraction factors $\gamma_{i/f} = \big[1 - (v_{i/f}/ c)^2\big]^{-1/2}$. The transition current density produced as the probing electron transitions between incoming $\psi_i({\bf x})=L^{-3/2}e^{i{\bf k}_i\cdot{\bf x}}$ and outgoing $\psi_f({\bf x})=L^{-3/2}e^{i{\bf k}_f\cdot{\bf x}}$ plane wave states is $\mathbf{J}_{fi}(\mathbf{x}) = -(e\hbar/2mL^3) (2\mathbf{k}_i - \mathbf{q})e^{i \mathbf{q}\cdot \mathbf{x}}$, while $\bar{\mathbf{G}}(\mathbf{x}, \mathbf{x}', \omega)$ is the dyadic Green's tensor characterizing the electromagnetic responses of the sample. Leveraging the plane wave form of $\mathbf{J}_{fi}(\mathbf{x})$, Eq. \eqref{DDCS_final_form} can be recast as 
\begin{equation}
\begin{split}
    \frac{\partial^2 \sigma}{\partial E_{if} \partial \Omega_f} &=  -\frac{1}{\pi}\Big(\frac{m v_i L^2}{2 \pi\hbar^2}\Big)^2 \bigg( \frac{ k_{f}}{k_i} \bigg) \sum_{\kappa\kappa'}\textrm{Im}\Big[ \mathbf{E}_{fi,{\bf q}\kappa}^{0*}(\varepsilon_{if}) \cdot \bar{\boldsymbol{\Pi}}_{\mathbf{q}\kappa\kappa'}(\varepsilon_{if}) \cdot \mathbf{E}^0_{fi,{\bf q}\kappa'}(\varepsilon_{if}) \Big]\\
    &=-\frac{1}{\pi}\Big(\frac{m v_i L^2}{2 \pi\hbar^2}\Big)^2 \bigg( \frac{ k_{f}}{k_i} \bigg) \sum_\kappa\textrm{Im}\Big[ \mathbf{E}_{fi,{\bf q}\kappa}^{0*}(\varepsilon_{if}) \cdot \mathbf{p}_{n0,{\bf q}\kappa}(\varepsilon_{if}) \Big] \\
    \end{split}
    \label{DDCS_q_space}
\end{equation}
where
\begin{equation}
    \mathbf{E}^0_{fi,\mathbf{q}\kappa}= \frac{2\pi  ie \gamma_i}{k_i L^2 \omega}  \frac{({\omega}/{c})^2\bar{\bf{I}} - \mathbf{q}\mathbf{q}}{({\omega}/{c})^2 - q^2}  \cdot(2 \mathbf{k}_i - \mathbf{q}) e^{i \mathbf{q} \cdot \mathbf{r}_\kappa}
    \label{vacuum_field}
\end{equation}
is the Fourier coefficient of the probe's vacuum transition field $\mathbf{E}^0_{fi}(\mathbf{x}, \omega) = - 4 \pi i \omega\int d{\bf x}' \, \bar{\mathbf{G}}_0(\mathbf{x}, \mathbf{x}', \omega) \cdot (L/v_i)\mathbf{J}_{fi}(\mathbf{x}')$ with free space Green's dyadic $\bar{\mathbf{G}}_0(\mathbf{x}, \mathbf{x}', \omega)=(-1/4\pi\omega^2)[(\omega/c)^2\bar{\bf I}+\nabla\nabla]{e^{i ({\omega}/{c}) |\mathbf{x} - \mathbf{x'}|}}/{|\mathbf{x} - \mathbf{x}'|}$. The Fourier coefficient is related to the position- and frequency-dependent field, evaluated at ${\bf x}=(\mathbf{x}_{\mathbf{n} \kappa},z=0)$ and $\omega = \varepsilon_{if}$, as $\mathbf{E}^0_{fi}({\bf x}_{{\bf n}\kappa}, \varepsilon_{if})\equiv\mathbf{E}^0_{fi,{\bf n}\kappa}(\varepsilon_{if})=\mathbf{E}^0_{fi,{\bf q}\kappa}e^{i{\bf q}\cdot{\bf x}_{\bf n}}$. Intrinsic excitations of the sample are encoded within the tensor-valued $\mathbf{q}$-dependent response function $\bar{\boldsymbol{\Pi}}_{\mathbf{q}\kappa\kappa'}(\varepsilon_{if}),$ while the subset that are excited by the electron probe are characterized by the induced moment ${\bf p}_{n0,{\bf q}\kappa}(\varepsilon_{if})=\sum_{\kappa'}\bar{\boldsymbol{\Pi}}_{\mathbf{q}\kappa\kappa'}(\varepsilon_{if})\cdot \mathbf{E}^0_{fi,{\bf q}\kappa'}(\varepsilon_{if})$.

Despite apparent physical distinctions, the discrete translation symmetry shared by each structure necessitates periodic energy-momentum dispersion of both phonon and LPP quasiparticles excitations. In either class of 2D material, Bloch wave excitations with in-plane momentum $\hbar\mathbf{K}$ can be described by the equations of motion (SI)
 \begin{equation} 
   -\omega^2 {\vec{u}}_{\mathbf{K} \kappa}-i\eta\omega{\vec{u}}_{\mathbf{K} \kappa} + \sum_{\kappa'}\bar{\boldsymbol{\cal D}}_{\kappa \kappa'}(\vec{K}, \omega)\cdot{\vec{u}}_{\mathbf{K} \kappa'} = Z_{\kappa}e M_{\kappa}^{-1} \mathbf{E}_{\mathbf{K} \kappa}^{0},
   \label{EOM}
\end{equation}
where the displacement of the coordinate at lattice position $\mathbf{x}_{\mathbf{n} \kappa}$ is $\mathbf{u}_{\mathbf{n} \kappa} = \mathbf{u}_{\mathbf{K} \kappa} e^{i \mathbf{K}\cdot \mathbf{x}_{\mathbf{n}}}e^{-i \omega t}$. Electron beams serve as one example of a probe of reciprocal space lattice excitations where the in plane recoil momentum $\hbar{\bf q}_\parallel$ is the Bloch momentum $\hbar{\bf K}$. However, propagating free space light can also drive lattice responses at specific ${\bf K}$ points corresponding to the projection ${\bf k}_\parallel$ of its wavevector ${\bf k}$ onto the sample plane. In either case, the induced moments are ${\bf p}_{n0,\mathbf{K}} = [Z_1e{\bf u}_{\mathbf{K}1}, \dots, Z_s e {\bf u}_{\mathbf{K}s}] = [(-\omega^2-i\eta \omega)\bar{\bar{\mathbf{I}}} +\bar{\bar{\boldsymbol{\cal D}}}(\vec{K},\omega)]^{-1} [(Z_1 e)^2\mathbf{E}_{{\bf K} 1}^{0} /M_1,\ldots,(Z_s e)^2\mathbf{E}_{{\bf K}s}^{0}/M_s] = \bar{\bar{\boldsymbol{\Pi}}}_{\bf K}(\omega)\mathbf{E}^0_\mathbf{K}$, where the $3s\times1$ vector $\mathbf{E}^0_\mathbf{K}=[\mathbf{E}_{{\bf K} 1}^{0},\ldots,\mathbf{E}_{{\bf K}s}^{0}]$ represents the specific probe under consideration evaluated at each of the $s$ sublattice sites. Double bars have been introduced to distinguish $3s \times 3s$ matrices from $3 \times 3$ matrices, which have a single bar, e.g., $\big[ \bar{\bar{\boldsymbol{\Pi}}}_{\mathbf{K}}(\omega) \big]_{\kappa \kappa'} = \bar{\boldsymbol{\Pi}}_{\mathbf{K}\kappa\kappa'}(\omega)$, while the absence of a $\kappa$ subscript indicates that a vector is $3s \times 1$.

\begin{figure}
    \centering
\includegraphics{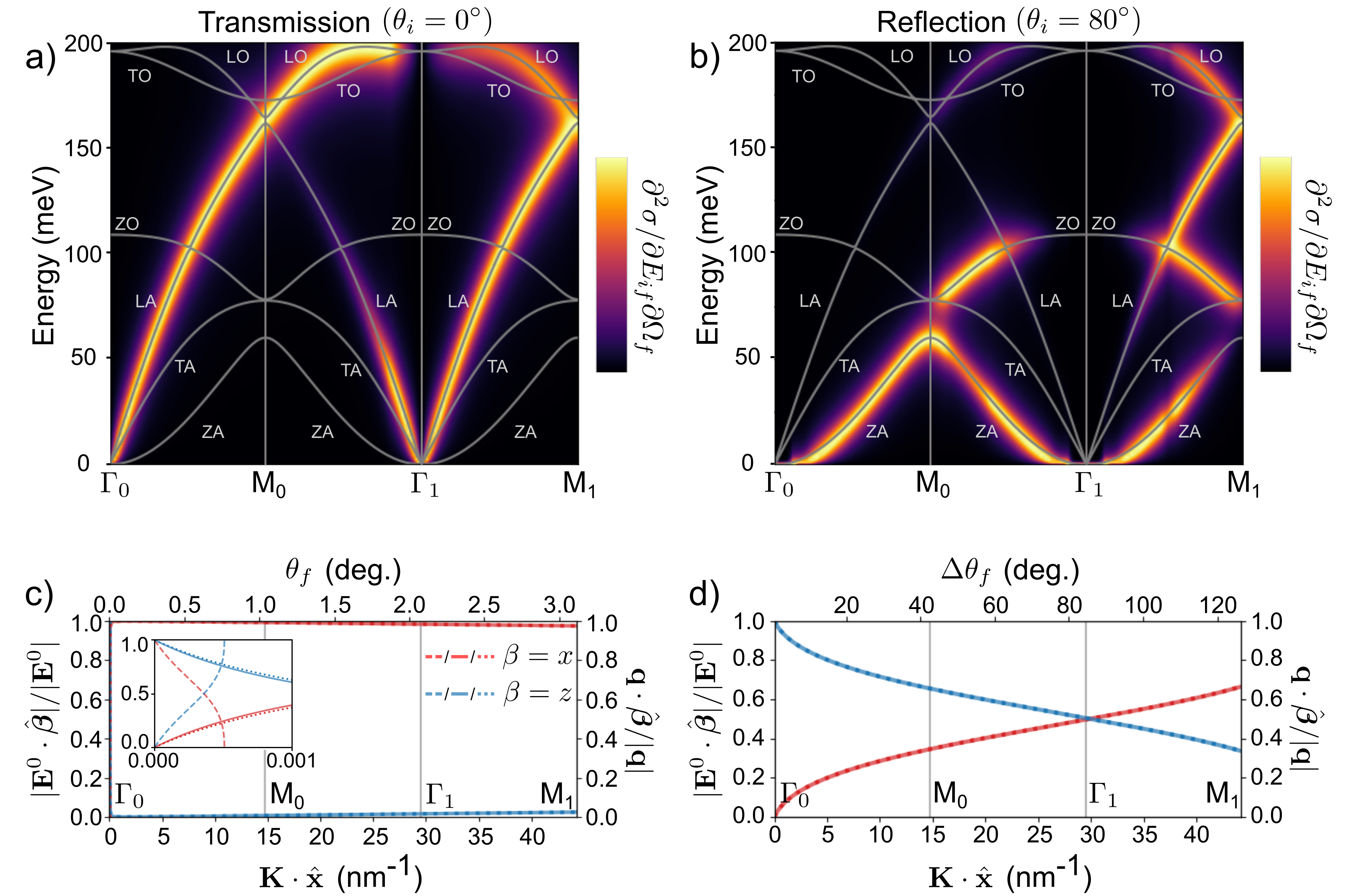}
\caption{Energy-momentum dispersion of graphene phonons probed along the $\Gamma-\textrm{M}$ direction. a) and b) display the $q$-EELS DDCS in the transmission and reflection geometries, respectively. Each spectrum at fixed ${\bf q}_\parallel$ is independently normalized to the maximum DDCS value at that ${\bf q}_\parallel$ point as in Ref.\cite{senga2019position}. Gray traces indicate the dispersion of the graphene phonon eigenenergies. c) and d) show the normalized magnitude of the in-plane $\mathbf{E}^{0}_{fi} \cdot \hat{\mathbf{x}}$ (red) and out-of-plane $\mathbf{E}^{0}_{fi} \cdot \hat{\mathbf{z}}$ (blue) components of the electron's vacuum transition field at $E_{if}=100$ meV versus $\vec{q}_\parallel$ along $\Gamma_0-\textrm{M}_1$, computed in the fully-retarded (solid) and quasistatic (dotted) limits in transmission (c) and reflection (d) geometries. Note the scale difference between $\theta_f$ and $\Delta \theta_f$ in (c) and (d). The inset in c) shows the dispersion of $\mathbf{E}^{0}_{fi}$ as well as that of free space plane wave light ${\bf E}^0=E_0\hat{\boldsymbol{\epsilon}}({\bf k})$ in the immediate vicinity of the $\Gamma_0$ point, where the dashed lines represent the in-plane $\mathbf{E}^{0} \cdot \hat{\mathbf{x}}$ (red) and out-of-plane $\mathbf{E}^{0} \cdot \hat{\mathbf{z}}$ (blue) components of the incident optical field at $\hbar\omega=100$ meV versus in-plane wave vector ${\bf k}_\parallel=(\vec{k}\cdot\hat{\vec{x}})\hat{\bf x}$. In a) and c), the incident electron wave vector $\mathbf{k}_i$ is oriented normal to the sample plane ($\theta_i=0^{\circ}$) with initial velocity $v_i = 0.3c$. In b) and d), $\theta_i=80^{\circ}$ with respect to the surface normal in the $xz$ scattering plane and $v_i = 0.01c$. All calculations include an empirical damping rate of $\eta=20$ $\textrm{meV}/\hbar$ at zero temperature and a vacuum background.}
\label{F2}
\end{figure}

Focusing first on atomic crystal phonons where the resonant IR wavelengths together with subnanometer bond lengths justify neglect of the finite speed of light, the generalized dynamical matrix $\bar{\boldsymbol{\cal D}}_{\kappa\kappa'}(\vec{K},\omega)$ reduces to the conventional dynamical matrix $\bar{ \mathbf{D}}_{\kappa\kappa'}(\vec{K})=\bar{\mathbf{F}}_{\kappa\kappa'}(\vec{K})+\bar{\mathbf{C}}_{\kappa\kappa'}(\vec{K})$ \cite{dove1993lattice}, which accounts for the sum of covalent $\bar{\mathbf{F}}_{\kappa\kappa'}(\vec{K})$ and ionic $\bar{\mathbf{C}}_{\kappa\kappa'}(\vec{K})$ interactions (SI). We compute all electron-induced graphene excitations within the harmonic crystal approximation \cite{dove1993lattice, hohenester2018inelastic} based upon Eq. \eqref{EOM} using an empirically-parameterized model \cite{bosak2007elasticity,michel2008theory} for $\bar{\mathbf{F}}_{\kappa\kappa'}(\mathbf{q}_{\parallel})$ and $\mathbf{q}$-dependent effective charges $Z_\kappa({\bf q}_\parallel)$ along $\Gamma-\textrm{M}$ from Ref. \cite{senga2019position} (SI). Within the sample and scattering planes lie the longitudinal acoustic (LA) and optical (LO) phonons, while the transverse acoustic (TA) and optical (TO) phonons are polarized within the sample plane but are oriented perpendicular to the scattering plane. Oscillations of the $z$-directed acoustic (ZA) and optical (ZO) phonons are directed normal to the sample plane. Figs. \ref{F2}a,b display the calculated energy-momentum dispersion $\hbar\omega_\lambda({\bf K}')$ of the phonon eigenmodes (gray traces), where $\lambda=1-6$ denotes the band indices, as well as the phononic responses induced under wide field (plane wave) electron excitation as encoded in the $q$-EELS DDCS collected in the transmission (panel a) and reflection (panel b) geometries, where momentum conservation requires $\mathbf{K} = \mathbf{q}_\parallel =({\bf q}\cdot \hat{\bf x})\hat{\bf x} = \mathbf{K}' + \mathbf{G}$, with $\mathbf{G}$ a reciprocal lattice vector. At each $\vec{q}_{\parallel}$ point, the spectrum is independently normalized to the maximum DDCS value, as in Ref. \cite{senga2019position}. For unnormalized $q$-EELS DDCS spectra, see Fig. S1. In Fig. \ref{F2}a, the incident plane wave wave vector $\mathbf{k}_i$ of the electron probe is normal to the graphene surface ($\theta_i=0^{\circ}$) and its initial speed is $v_i = 0.3c$, which are parameters typical of $q$-EELS experiments performed in the transmission geometry \cite{senga2019position,hong2020probing}. Reflection measurements, on the other hand, generally involve momentum transfers with a larger out-of-plane component (Figs. \ref{F1}a,b) such that the DDCS is largest for low incident electron kinetic energies and large angles of incidence $\theta_i$ \cite{politano2015emergence,politano2017spectroscopic,li2023direct}. Thus, for the reflection calculations in Fig. \ref{F2}b, we choose $\theta_i=80^{\circ}$ from the surface normal and $v_i = 0.01c$, typical of HREELS experiments. 


It is evident from Eq. \eqref{DDCS_q_space} that the selection rule for electron beam excitation of a specific phonon mode ${\boldsymbol{\xi}}_\lambda$ at ${\bf q}_\parallel$ of energy $\hbar\varepsilon_{if}$ requires overlap of the electron's vacuum field $\mathbf{E}^{0*}_{fi, \mathbf{q}}$ in Eq. \eqref{vacuum_field} with the induced phonon moment ${\bf p}_{n0}$ dominated by ${\boldsymbol{\xi}}_\lambda$ at ${\bf q}_\parallel$ and $\hbar\varepsilon_{if}$. However, in the quasistatic ($c\to\infty$) and dipole ($a\ll 2\pi c/\varepsilon_{if}$) limits relevant to graphene, the electron's transition field  $\mathbf{E}^0_{fi, \mathbf{q}}(\mathbf{x})\propto q^{-2}{\bf q}$, and the general induced phonon moments become the induced dipole moments, i.e., ${\bf p}_{n0}\to{\bf d}_{n0}$, so that the selection rule $\mathbf{E}^{0*}_{fi, \mathbf{q}}\cdot{\bf p}_{n0, \mathbf{q}}$ reduces to $\sum_{\kappa}{\bf q}\cdot{\bf d}_{n0, \mathbf{q} \kappa}=\sum_{\kappa} {\bf q}_\parallel\cdot{\bf d}_{n0, \mathbf{q} \kappa}^\parallel+{q}_z{d}_{n0, \mathbf{q} \kappa}^z$ when projected onto the sample coordinates. Based upon the dipole limit of the scattering form factor of a single Coulombically bound target electron, $\langle n|e^{i{\bf q}\cdot{\bf x}}|0\rangle\approx\langle n|1+i{\bf q}\cdot{\bf x}+\cdots|0\rangle\propto{\bf q}\cdot{\bf d}_{n0}$ \cite{SakuraiModern}, this result is expected as electromagnetic retardation effects are negligible for graphene. The $\sum_{\kappa}{\bf q}\cdot{\bf d}_{n0, \mathbf{q} \kappa}$ selection rule is also identical to that invoked to rationalize HREELS and $q$-EELS phonon measurements \cite{de2015symmetries, nicholls2019theory}.

While the $\sum_{\kappa}{\bf q}\cdot{\bf d}_{n0, \mathbf{q} \kappa}$ selection rule explains the absence of the TO and TA bands in the spectra presented in Fig. \ref{F2} on grounds of symmetry, the degree to which accessible bands contribute to the DDCS spectra inside (and outside) the first BZ can be more clearly understood by considering the top line of Eq. \eqref{DDCS_q_space}. In particular, the DDCS is proportional to $|\mathbf{E}_{fi, \mathbf{q}}^{0}|^2\,\textrm{Im}\Big[ \hat{\mathbf{E}}_{fi,\mathbf{q}}^{0*} \cdot \bar{\bar{\boldsymbol{\Pi}}}_{\mathbf{q}} \cdot \hat{\mathbf{E}}^0_{fi, \mathbf{q}} \Big]$, showing that the characteristics of the transition field largely dictate how strongly each tensor component of $\bar{\bar{\boldsymbol{\Pi}}}_{\mathbf{q}}$ contributes to the DDCS. Although the eigenmode dispersion is periodic in reciprocal space, the momentum dependence of the effective charge $Z(\mathbf{q})$ causes $\bar{\bar{\boldsymbol{\Pi}}}_{\mathbf{K}' + \mathbf{G}} \ne \bar{\bar{\boldsymbol{\Pi}}}_{\mathbf{K}'}$. The dispersions in ${\bf q}_\parallel$ of the electron's field $\mathbf{E}^0_{fi}$ (solid red/blue traces) as well as that of its limiting quasistatic form $\mathbf{E}^0_{fi}\propto\bf q$ (dotted red/blue traces) are displayed in Figs. \ref{F2}c,d for both transmission and reflection geometries, clearly showing the reduction of the former to the latter in the quasistatic limit appropriate to graphene. Notably, in transmission, the probe is strongly $x$-polarized except in the immediate vicinity of $\Gamma_0$, explaining the preferential excitation of both LA and LO phonons at both small and large momenta, as recently observed in $q$-EELS measurements of individual freestanding graphene monolayers \cite{senga2019position}.

In contrast, the probe's polarization in the reflection geometry (Fig. \ref{F2}d) evolves from predominantly $z$-polarized in the first BZ to predominantly $x$-polarized in the second BZ. This crossover in polarization content of the electron's field elucidates the computed DDCS in Fig. \ref{F2}b, where primary excitation of ZA and ZO phonons at low momenta \cite{politano2015emergence,politano2017spectroscopic} give way to the additional excitation of both LA and LO phonons outside of the first BZ. Taken together, these results contrast the different information contained within $q$-EELS measurements of graphene performed in transmission or reflection geometries at IR energies in the first two BZs. For comparison, optical excitation via propagating free space light is strongly limited in the IR by the light cone boundary, where ${\bf K}=|\mathbf{k}| \hat{\mathbf{x}} \sim10^{-3}$ nm$^{-1}$ (see dashed red/blue traces in the inset of Fig. \ref{F2}c).
 
\begin{figure}
\centering
\includegraphics{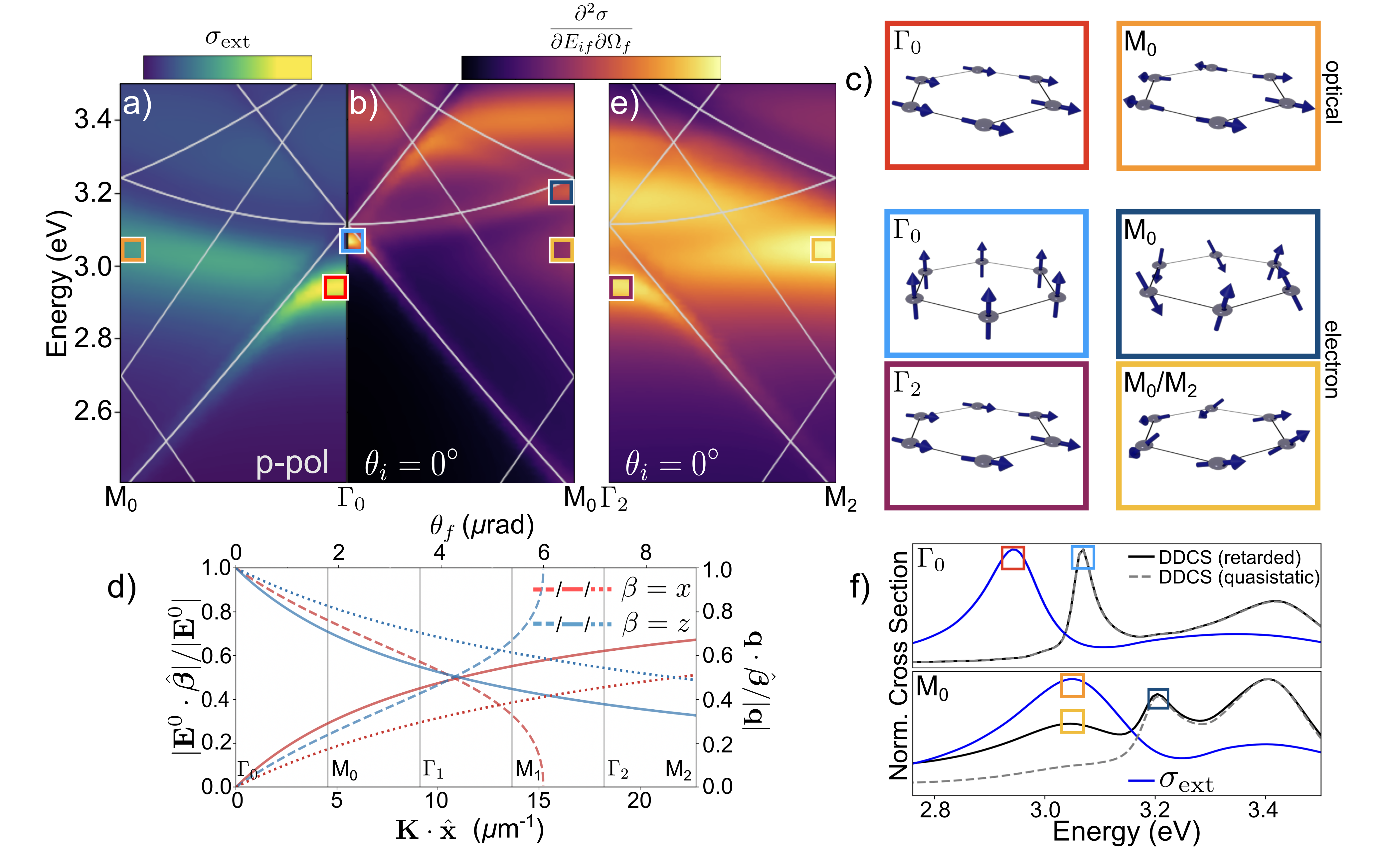}
\caption{Energy-momentum dispersion of LPPs along the $\Gamma-\textrm{M}$ direction in a honeycomb plasmonic array composed of 100 nm wide $\times$ 60 nm high silver nanodisks separated by 460 nm. The localized surface plasmon energies of each isolated nanodisk are 3.13 eV and 3.36 eV in the $x$ and $z$ directions, respectively. a) Normalized optical extinction cross section under $p$-polarized excitation as a function of incident wave vector ${\bf k}_\parallel$ in the first BZ. b) Normalized $q$-EELS DDCS calculated in the transmission geometry versus recoil wave vector ${\bf q}_\parallel$ in the first BZ. The electron's incident wave vector ${\bf k}_i$ is oriented normal to the sample surface with an initial velocity of $v_i=0.7c$. c) Induced LPP polarizations extracted at the boxed points in the optical extinction and $q$-EELS DDCS spectra. d) Polarization dispersion of the electron's vacuum transition field $\mathbf{E}^{0}_{fi}$ (solid red/blue) and of the free space plane wave optical field $\mathbf{E}^{0}$ (dashed red/blue) contrasted against that of the electron's vacuum field $\mathbf{E}^{0}_{fi}\propto{\bf q}$ in the quasistatic limit (dotted red/blue) along the $\Gamma_0-\textrm{M}_2$ direction, all computed at 3 eV. e) $q$-EELS DDCS under same conditions as b) along $\Gamma_2-\textrm{M}_2$. The white traces in (a), (b), and (e) are the photonic dispersion of the empty honeycomb lattice. f) Normalized momentum-resolved extinction (blue traces) and retarded/quasistatic DDCS spectra (black/gray-dashed traces) at $\Gamma_0$ and M$_0$ points from (a) and (b). All calculations use a vacuum background.}
\label{F3}
\end{figure}

Sharing the same hexagonal lattice structure as graphene but with a lattice constant of $a=460$ nm, Fig. \ref{F3} considers a 2D plasmonic honeycomb array composed of 100 nm diameter $\times$ 60 nm high silver nanodisks. In contrast to graphene, the large lattice constant and optical excitation frequencies of the array renders the entire BZ accessible to optical excitation \cite{juarez2022m}. Displayed in Fig. \ref{F3}a is the angle-resolved optical extinction cross section $\sigma_{\textrm{ext}}({\bf k},\omega) = (4 \pi \omega/c) \textrm{Im} \big[ \sum_\kappa \mathbf{E}^{0*}_{\mathbf{k} \kappa} \cdot \mathbf{p}_{n0, \mathbf{k}\kappa} \big],$ where $\mathbf{E}^{0}_{\mathbf{k}\kappa}=E_0\hat{\boldsymbol{\epsilon}}({\bf k})e^{i{\bf k}\cdot{\bf r}_\kappa}$ is the Fourier coefficient of the stimulating optical plane wave field with wave vector $\mathbf{k}$, while the fully retarded $q$-EELS DDCS (Eq. \eqref{DDCS_q_space}) is shown in Fig. \ref{F3}b, both within the first BZ. The inelastic electron scattering spectra are computed in the forward scattering geometry with ${\bf k}_i$ oriented along the $z$-axis ($\theta_i=0^\circ$) and $v_i=0.7c$. Both optical and electron beam LPP spectra are calculated from Eq. (\ref{EOM}) using the method of coupled dipoles where the multipolar plasmon moments ${\bf p}_{n0}$ in each silver nanodisk are approximated by individual point electric dipole moments ${\bf d}_{n0}$ \cite{bourgeois2021lattice,bourgeois2022optical} with anisotropic polarizability evaluated using tabulated dielectric data for silver \cite{PhysRevB.6.4370} (SI). Unlike for graphene, site-to-site coupling in nanoparticle arrays is mediated by long-range, causal interactions between nanophotonic antenna modes localized to each nanoscale scattering element. As a consequence, the generalized dynamical matrix $\bar{\boldsymbol{\cal D}}_{\kappa\kappa'}({\bf K},\omega)$ for LPPs encodes the individual nanoantenna resonances as well as the retarded dipole-dipole interactions between sites (SI). The white traces in Figs. \ref{F3}a,b denote the empty lattice photon dispersion in the reduced zone scheme \cite{guo2017geometry}. 
 
The honeycomb lattice LPPs are photonic analogs of the acoustic (in-phase) and optical (out-of-phase) transverse (TA, TO), longitudinal (LA, LO), and $z$-directed (ZA, ZO) phonon modes of graphene. Under $p$-polarization, optical extinction (Fig. \ref{F3}a) reveals the in-plane ($x$-polarized) in-phase LPP at $\Gamma_0$, while the electron DDCS (Fig. \ref{F3}b) probes the complementary out-of-plane ($z$-polarized) in-phase LPP at the same reciprocal space point, as seen by the color-coded induced moment plots in Fig. \ref{F3}c. The induced dipoles exhibit the expected in-phase and out-of-phase polarizations between sites within the unit cell at the high-symmetry points shown in Figs. \ref{F3}a,b. The complementary behavior exhibited by photon and electron probes in Figs. \ref{F3}a,b can again be understood by examining the dispersion of the electron's vacuum transition field $\mathbf{E}^{0}_{fi}$ (solid blue/red traces) versus that of plane wave light ${\bf E}^0$ (solid black/gray traces) shown in Fig. \ref{F3}d. In moving from the first to second BZ, the $p$-polarized optical field ${\bf E}^0$ switches from dominantly $x$-polarized to dominantly $z$-polarized at the light cone edge near M$_1$, while $\mathbf{E}^{0}_{fi}$ from Eq. \eqref{vacuum_field} is predominantly polarized in the $z$ direction in the first BZ, but evolves to be largely $x$-polarized at larger ${\bf K}=\mathbf{q}_{\parallel}$. Indeed the computed optical extinction spectrum in Fig. \ref{F3}a shows clear signatures of the $x$-polarized in-phase and out-of-phase LPPs with increasing ${\bf k}_\parallel$ in the first BZ. Near the light cone edge, the $z$-polarized in-phase LPP becomes evident in $\sigma_{\textrm{ext}}({\bf k},\omega)$, however, the $z$-polarized out-of-phase LPP is not apparent (Fig. S2a). The evolution of $\sigma_{\textrm{ext}}({\bf k},\omega)$ under $s$- and $p$-polarized excitation as $\mathbf{k}_{\parallel}$ traverses the full boundary of the irreducible BZ is shown in Fig. S3a,b. Oppositely, but reflecting the crossover behavior in $\mathbf{E}^0_{fi}$ polarization at larger ${\bf q}_\parallel$ values in Fig. \ref{F3}d, the electron probe excites primarily the $z$-polarized out-of-phase LPP at M$_0$ as well as both $x$-polarized out-of-phase and in-phase LPPs at M$_0$/M$_2$ and $\Gamma_2$, respectively, in the third BZ (Fig. \ref{F3}e), a region inaccessible to free space optical excitation. The $\mathbf{q}_\parallel$ value where $x$- and $z$-polarized $\mathbf{E}^0_{fi}$ cross over can be tuned across multiple BZs in the transmission geometry by varying the primary kinetic energy of the probing electron (Fig. S4). In the reflection geometry typical of HREELS, and in contrast to optical excitation, the electron's transition field is strongly $z$-polarized across many BZs, which limits the induced sample polarizations to be predominantly $z$-polarized (Fig. S5).

In contrast to the near indistinguishability of quasistatic and fully-retarded transition fields under the conditions presented in Figs. \ref{F2}c,d for graphene, Fig. \ref{F3}d displays noticeable differences in $\mathbf{E}_{fi}^0$ arising from the finite speed of light. The consequences of these differences are evident in Fig. \ref{F3}f, which displays normalized momentum-resolved optical (blue traces) and electron beam spectra at $\Gamma_0$ and M$_0$, where both fully retarded (black traces) and quasistatic (dashed gray traces) descriptions of the electron's vacuum transition field are adopted. While the normalized DDCS spectra are indistinguishable at the $\Gamma_0$ point, where the quasistatic and fully-retarded fields in Fig. \ref{F3}d coincide, it is evident from the M$_0$ point spectra that a quasistatic description of the electron's field fails to fully capture the electron's ability to excite the $x$-polarized out-of-phase LPP in the first BZ at $\sim3.05$ eV. As compared to the quasistatic case, a fully retarded description of the electron's field involves a larger ratio of $x$- to $z$-polarized $\mathbf{E}^{0}_{fi}$ components, leading to stronger excitation of the $x$-polarized out-of-phase LPP. 


While of considerable utility in the investigation of 2D atomic materials, spectroscopies derived from the inelastic scattering of free electron beams have found less consideration as probes of 2D periodic nanostructures. Here, we have presented a fully-retarded theoretical approach to describe the low energy inelastic scattering of wide field electron beams from 2D periodic materials spanning from atomic-scale quantum materials to nanophotonic array structures. In both forward and reflection geometries typical of $q$-EELS and HREELS measurements, we have investigated the reciprocal-space excitations native to a plasmonic honeycomb lattice under both inelastic electron scattering and angle-resolved optical extinction spectroscopies, and compared these observables to the momentum-resolved phonon excitation spectrum of monolayer graphene. We find important differences arise in the selection rules between 2D atomic and nanophotonic materials under optical- and electron-based probes, leading to complementary behavior in their polarization-dependent responses both inside and outside of the BZ. These differences, together with a breakdown of the quasistatic approximation, thus require consideration of fully retarded light-matter interactions to properly interpret inelastic electron scattering signals from periodic nanophotonic materials.

\begin{acknowledgement}
All work was supported by the U.S. Department of Energy (DOE), Office of Science, Office of Basic Energy Sciences (BES), Materials Sciences and Engineering Division under Award No. DOE BES DE-SC0022921.
\end{acknowledgement}

\begin{suppinfo}
Additional details related to the employed graphene phonon model, two-dimensional lattice plasmon polariton response matrix, additional graphene and lattice plasmon dispersion diagrams, and the incident electron speed dependence of the transition field.
\end{suppinfo}

\bibliography{manuscript}

\end{document}